\documentclass[conference]{IEEEtran}
\IEEEoverridecommandlockouts
\usepackage{cite}
\usepackage{amsmath,amssymb,amsfonts}
\usepackage{algorithmic}
\usepackage{graphicx}
\usepackage{textcomp}
\usepackage{hyperref}
\usepackage{url}
\usepackage{xcolor}
\usepackage[]{algorithm2e}
\usepackage{subcaption}

\def\BibTeX{{\rm B\kern-.05em{\sc i\kern-.025em b}\kern-.08em
    T\kern-.1667em\lower.7ex\hbox{E}\kern-.125emX}}
\begin{document}

\title{TempCharBERT: Keystroke Dynamics for Continuous Access Control Based on Pre-trained Language Models}

\author{\IEEEauthorblockN{\textit{Matheus Simão\textsuperscript{4}, Fabiano Prado\textsuperscript{4}, Omar Abdul Wahab\textsuperscript{3}, Anderson Avila\textsuperscript{1,2}}}\\
\IEEEauthorblockA{\textsuperscript{1}Institut national de la recherche scientifique (INRS-EMT), Université du Québec, Montréal, Québec, Canada  \\
\textsuperscript{2}INRS-UQO Mixed Research Unit on Cybersecurity, Gatineau, Québec, Canada \\
\textsuperscript{3}Department of Computer and Software Engineering, Polytechnique Montréal, Montréal, Québec, Canada}
\textsuperscript{4}São Paulo State Technological College, São Paulo, Brazil\\
}
\maketitle
\begin{abstract}
With the widespread of digital environments, reliable authentication and continuous access control has become crucial. It can minimize cyber attacks and prevent frauds, specially those associated with identity theft. A particular interest lies on keystroke dynamics (KD), which refers to the task of recognizing individuals' identity based on their unique typing style. In this work, we propose the use of pre-trained language models (PLMs) to recognize such patterns. Although PLMs have shown high performance on multiple NLP benchmarks, the use of these models on specific tasks requires customization. BERT and RoBERTa, for instance, rely on subword tokenization, and they cannot be directly applied to KD, which requires temporal-character information to recognize users. Recent character-aware PLMs are able to process both subwords and character-level information and can be an alternative solution. Notwithstanding, they are still not suitable to be directly fine-tuned for KD as they are not optimized to account for user's temporal typing information (e.g., hold time and flight time). To overcome this limitation, we propose TempCharBERT, an architecture that incorporates temporal-character information in the embedding layer of CharBERT. This allows modeling keystroke dynamics for the purpose of user identification and authentication. Our results show a significant improvement with this customization. We also showed the feasibility of training TempCharBERT on a federated learning settings in order to foster data privacy.
\end{abstract}


\section{Introduction}

\begin{figure}
    \centering
    \includegraphics[width=.95\linewidth]{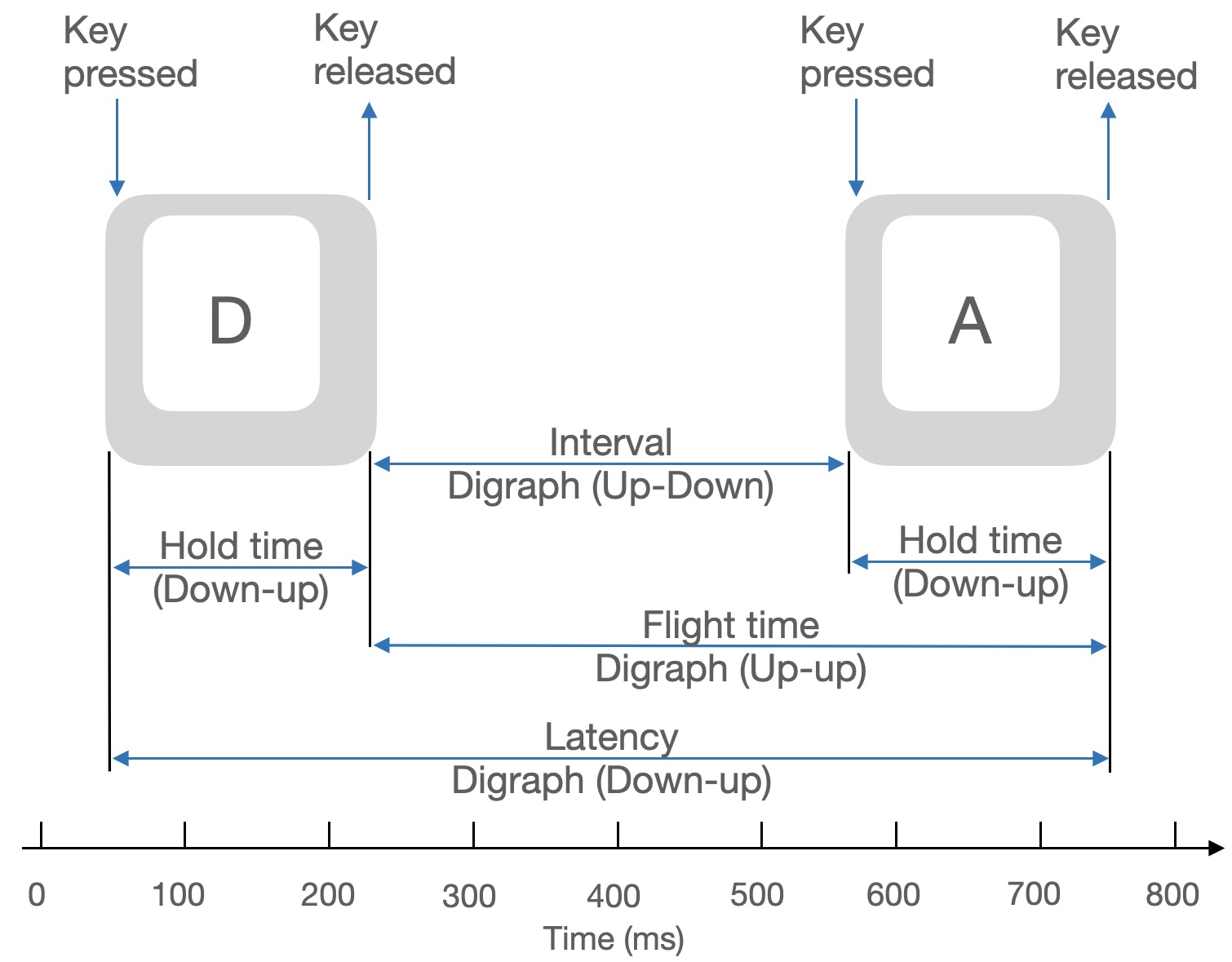}
    \caption{Keystroke metrics based on pressing and releasing timestamps, including latency interval, dwell (or hold) time and flight time.}
    \label{fig:keystroke_fea}
\end{figure}
In this work, we propose TempCharBERT, an architecture based on pre-trained language models (PLMs) and customized for KD. Although PLMs have been providing impressive results on multiple NLP benchmarks, the use of such models for specific tasks is not always straightforward and often requires some level of customization. One particular concern, for instance, is the tokenization granularity, which can be language sensitive, dependent of resource availability and can ultimately impact the performance of downstream tasks \cite{mielke2021between}. BERT and RoBERTa-based models, for example, use wordpiece tokenization and focus on processing subword inputs, rather than word or character-level tokens. This is not suitable for specific tasks, such as KD, which relies significantly on processing temporal information from each key used while typing. Thus, these models provide low performance if directly applied to keystroke dynamics. Recent character-aware PLMs, such as CharBERT \cite{ma2020charbert} and CharacterBERT \cite{boukkouri2020characterbert}, offer an alternative as they can process both subwords and character-level information. These models, however, are still not suitable to be directly fine-tuned for keystroke dynamics as they are not optimized to account for user's temporal typing information (e.g., dwell time and flight time\footnote{The concepts of dwell time and flight time are explained in section 2.}). To address this limitation, we propose a modification on the CharBERT architecture. This change aims at incorporating keystroke metrics into the embedding layer of CharBERT. We show that these temporal-character information are enough to enhance the representation of user typing pattern, leading to significant improvement in terms of accuracy for user identification as well as in terms of equal error rate (EER) for user authentication. We also investigated the feasibility of training the proposed TempCharBERT in the Federated Learning settings to foster user data privacy. We found a small decay in the performance when compared to the centralized training settings. Thus, our contribution is summarized as follow:

\begin{itemize}
    \item We propose TempCharBERT as a new variant of CharBERT for keystroke dynamics. The originality of TempCharBERT stems from the temporal typing dynamic, which is crucial for capturing typing style.
    \item We evaluate the proposed model on two important tasks: user identification and user authentication. Our results suggest a significant improvement in terms of accuracy and equal error rate compared to the pure CharBERT and other baseline approaches.
    \item We show that the representation based on our customized embedding layer carries out meaningful user discriminative information.
    \item We also show that the representation attained can be successfully used on other architectures, such as the Long-Short term Memory (LSTM).
    \item To foster privacy, we show the feasibility of training TempCharBERT on the Federated Learning (FL) settings.
\end{itemize}

The remainder of this paper is organized as follows. Section 2 presents background material and related works. Section 3 describes the proposed method, while Section 4 presents the experimental setup. Section 5 presents the experimental results. Section 6 discuss the paper limitation and Section 7 concludes the paper with final considerations.

\section{Background and Related Work}
\subsection{Keystroke Dynamics}

Keystroke dynamics is a technology that allows the recognition of individuals' identity based on their typing style. It is achieved via measuring the typing rhythm of someone using a computer keyboard, touch-pad or a mobile device touchscreen, to name a few \cite{roy2022systematic}. As a behavioral biometrics, it can analyze not just what is being typed but most importantly the way it is being typed. Similarly to written signatures that are influenced by neural-physiological traits, while typing on a keyboard, a person leaves her or his digital signature as keystroke latencies (i.e., the time elapsed between keystrokes) \cite{joyce1990identity}.


KD can be applied during the authentication processes, right before starting the user new session. That is, when the user types her or his credential to be logged-in, a keystroke authentication system takes place to extract the temporal features described above and check if these features match the user template attained during the enrollment process. This is based on predefined arrangement, typically based on fixed-text, and is also referred to as \textit{static mode} \cite{roy2022systematic}. If the patterns are different, the access is denied, even if the \textit{login} and \textit{password} provided are correct, characterizing a two-factor authentication scheme. This prevents intruders that may have had access to users' credentials to unauthorizedly log-into a system. After the initial authentication, KD can still be useful as a continuous access control system that monitors the keyboard after the user session has been initialized, enabling behavioral intrusion detection \cite{zanero2004behavioral}. This is based on free-text and is referred to as \textit{dynamic mode} \cite{roy2022systematic}.

Recently, we have witnessed an attempt to adapt Transformers based models to be used on keystroke dynamics. Next, we discuss these approaches and how our proposed solution relates to them.

\subsection{Keystroke Dynamics based on Transformers}

Since the introduction of Transformers in \cite{vaswani2017attention}, the architecture has become the main building block for language representation \cite{devlin2018bert} and generation \cite{radford2018improving}, allowing many breakthroughs in natural language processing (NLP), as well as in other domains \cite{child2019generating}. To date, the use of Transformers for keystroke dynamics is almost inexistent, with only a handful of works in the literature. In \cite{stragapede2023mobile}, for instance, the authors introduced the innovative utilization of the Transformer architecture to construct a KD model. The authors proposed few changes on the traditional Transformer architecture. First, they replaced the positional encoding by the Gaussian range encoding presented in \cite{li2021two}. A second change made by the authors was to replace the original vanilla layer of Transformers by two modules: a Temporal Module and a Channel Module. The Temporal Module is responsible to extract information from the original input sequence, i.e., temporal-over-channel features, and the Channel Module extracts channel-over-temporal features after transposing the input sequence. Each module contains its own stack of identical layers with a multi-head self-attention mechanism and a multi-scale keystroke CNN. To the best of our knowledge, this work stands out as one of the only that has explored Transformers for KD. 

\begin{figure*}
    \centering
    \includegraphics[width=1\linewidth]{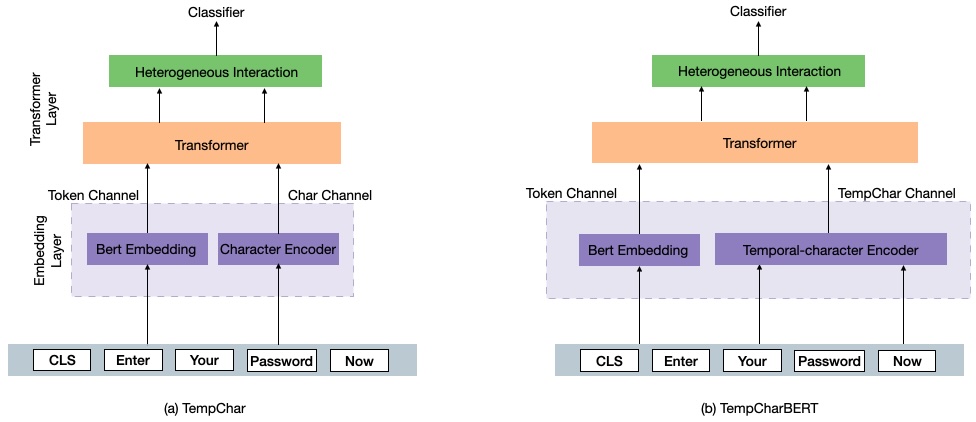}
    \caption{Comparison of the contextual word representation in CharBERT and the TempCharBERT architecture with the proposed Temporal-character Encoder for keystroke dynamics.}
    \label{fig:architecture_diff}
\end{figure*}

\subsection{Character-aware Language Models}
To the best of our knowledge, PLMs haven't yet been applied to KD, with the main challenge being the granularity of the input tokens \cite{mielke2021between}. While most PLMs build word representations from subword tokenization \cite{ma2020charbert}, KD relies on temporal-character information. Recently, the NLP community has started to address some of the issues of models trained based on standard algorithms for subword tokenization, such as Byte Pair Enconding \cite{sennrich-etal-2016-neural} and Wordpiece \cite{wu2016google}. That is because rigid tokenization makes it harder for models to cope with morphological and compositional variation in language \cite{tay2021charformer}. It also presents as weakness the inability to handle rare and novel words that were rarely seen during training, or out of vocabulary (OOV) words \cite{mielke2021between}. To this end, different initiatives have emerged to improve the ability of models to deal with rare and novel words. These are fundamentally word-based approaches that incorporates handling characters that make up a word \cite{mielke2021between}. This gives PLMs the capability not just to process words not seen during training, but also to be fine-tuned on downstream tasks that relies on character information, such as keystroke dynamics.

The authors in \cite{el-boukkouri-etal-2020-characterbert} proposed a variant of BERT, namely CharacterBERT. The model replace the wordpiece system of BERT by a Character-CNN module that represents entire words by consulting their characters. Instead of splitting unknown tokens into multiple wordpieces and then using a wordpiece embedding matrix to embed each unit independently, CharacterBERT uses the Character-CNN module to retrieve the characters of a token to produce a single representation. The authors showed that this change led to improved performance and more robust, word-level, and open-vocabulary representations. Similarly, in \cite{ma2020charbert}, the authors proposed the character-aware PLM named CharBERT. It is based on two encoding layers, referred to as token embedding and characher embedding. The former is a vanilla BERT embedding layer that provides token embedding based on subwords and the latter is a character encoder that provides token-level embeddings from a sequence of characters. Both embeddings are sent to the Transformer layer. Token and character representations are then fused and split by the heterogeneous interaction module after undergoing each Transformer layer. In a more recent work \cite{tay2021charformer}, the authors proposed the CharFormer. The model learns a subword tokenization during training based on the soft gradient-based subword tokenization module (GBST). This module can automatically learn latent subword representations from characters based on the training data. 


\section{TempCharBERT}

In this section, we present our proposed solution for keystroke dynamics named TempCharBERT. We start by describing the general architecture in Section 3.1. In Section 3.2, we give details on the temporal-character encoder and a comparative analysis between CharBERT and TempCharBERT embeddings is provided in Section 3.3  

\subsection{Model Architecture}
The proposed TemCharBERT has the same general architecture as CharBERT. However, as can be seen in Figure \ref{fig:architecture_diff}, an adaptation is required in order to incorporate user keystroke temporal information into the model. Thus, given a sequence of subwords $\{w_1, w_2, ..., w_m\}$ of length $m_i$, the characters from subword $w_i$ is represented by $\{c^i_1,... c^i_{n_i} \}$ and similarly the temporal information from keystroke timestamps is defined as $\{t^i_1,... t^i_{n_i}\}$. Note that $n_i$ is the number of characters in $w_i$ and that the the total number of characters and timestamps in a sentence is the same and given by $\sum^m_{i=1} n_i$. The main difference between the two architectures is in the Embedding layer. While CharBERT models the input words encoding only subwords and character information, TempCharBERT utilizes its Temporal-character Encoder to encode temporal information associated with each keystroke. Thus, TempCharBERT is still have two channels as CharBERT, but with the difference that the \textit{Char Channel} now comprises temporal-character information and is referred to as \textit{TempChar Channel}. Note that the proposed change in the embedding layer has only direct impact on the character representation. However, as the information propagates into the Transformer layers and specially into the Heterogeneous Interaction layers, it will also affect the overall representation.

\subsection{Temporal Encoder}

The proposed method enriches the character representation of CharBERT by incorporating in the Embedding Layer temporal information from users' keystrokes. Note that one of the key ideas of the CharBERT architecture is to create more robust representations capable of dealing with new words as well as with noise and typos in senteces. This is achieved by constructing contextual word representations from a sequence of character ones. The first step in this process is to create a fixed-size vector for each character as described below:

\begin{equation}
    e^i_j = W_c*c^i_j
\end{equation}

\noindent where $W_c$ is the character embedding matrix. The contextual word representation is attained from a bi-directional recurrent neural network, referred to as bi-GRU, as defined below:

\begin{equation}
    h^i_j(x) = \textbf{BI-GRU}(e^i_j)
\end{equation}

\noindent with $h^i_j(x)$ being the hidden state of the $j$-th character in the $i$-th token. The token-level embedding from characters is attained by concatenating the first and last hidden state and is represented as:

\begin{equation}
    h^i(x) = [h^i_1(x);h^i_{n_i}(x)]
\end{equation}

\noindent where $h^i(x)$ captures the full word information even for new words. 

The first step to adapt CharBERT is to create a fixed-size embedding of the keystroke timestamps. The two temporal information adopted in this work are dwell time and flight time, represented here by $d^i_j$ and $f^i_j$, respectively, for the $j$-th character in the $i$-th token. The temporal information is encoded as follow:

\begin{figure}
    \centering
    \includegraphics[width=.89\linewidth]{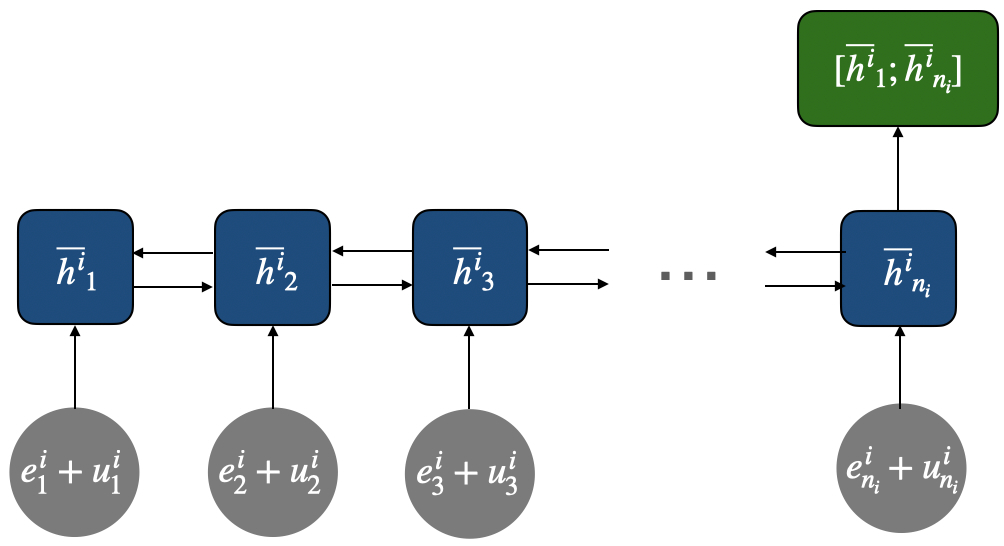}
    \caption{Temporal-character Encoder that captures full word information from characters that comprise temporal information.}
    \label{fig:tempcharenc}
\end{figure}

\begin{equation}
    u^i_j = T_c*d^i_j + T_c*f^i_j
\end{equation}

\noindent where $T_c$ is the temporal matrix embedding and $u^i_j$ represents the temporal embedding for the $i$-th token and $j$-th character. Note that to attain the contextual word representation with temporal-character information, both $e^i_j$ and $u^i_j$ are now passed through the bi-GRU, as shown below: 

\begin{equation}
    \overline{h^i}_j(x) = \textbf{BI-GRU}(e^i_j + u^i_j)
\end{equation}

\noindent with $\overline{h^i}_j(x)$ being the hidden state of the $j$-th character in the $i$-th token, now adapted to carry out also temporal information. The token-level embedding from temporal-characters is attained by concatenating the first and last hidden states and is represented as:

\begin{equation}
    \overline{h^i}(x) = [\overline{h^i}_1(x);\overline{h^i}_{n_i}(x)]
\end{equation}

\noindent where $\overline{h^i}(x)$ is the new hidden state representing the full word information together with the temporal cues for a given user. Figure \ref{fig:tempcharenc} shows how the contextual word representation based on temporal character information is obtained.

\subsection{Temporal Embeddings}

\begin{figure*}%
    \centering
    \subfloat{{\includegraphics[width=0.42\linewidth]{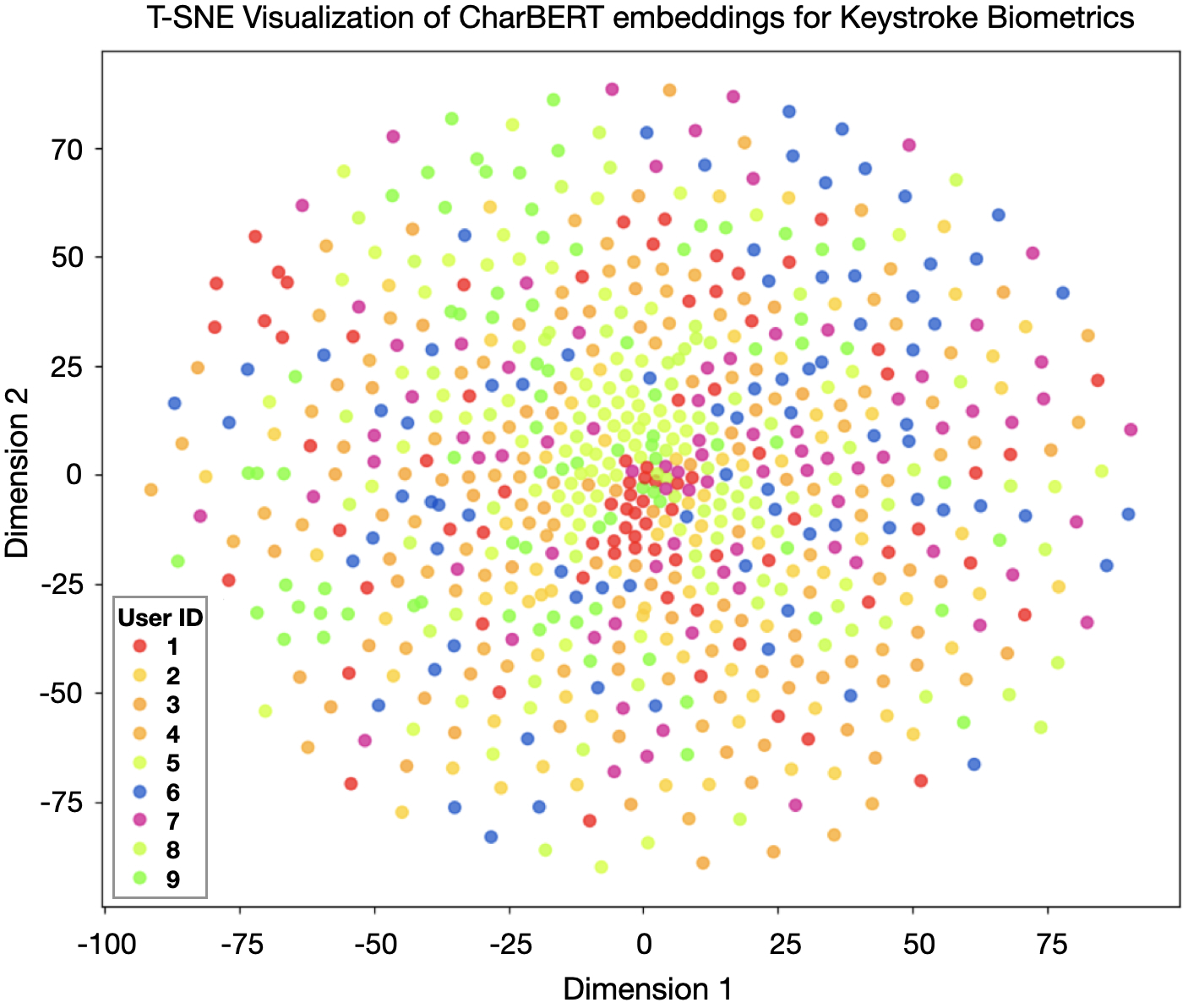} }}%
    \qquad
    \subfloat{{\includegraphics[width=0.42\linewidth]{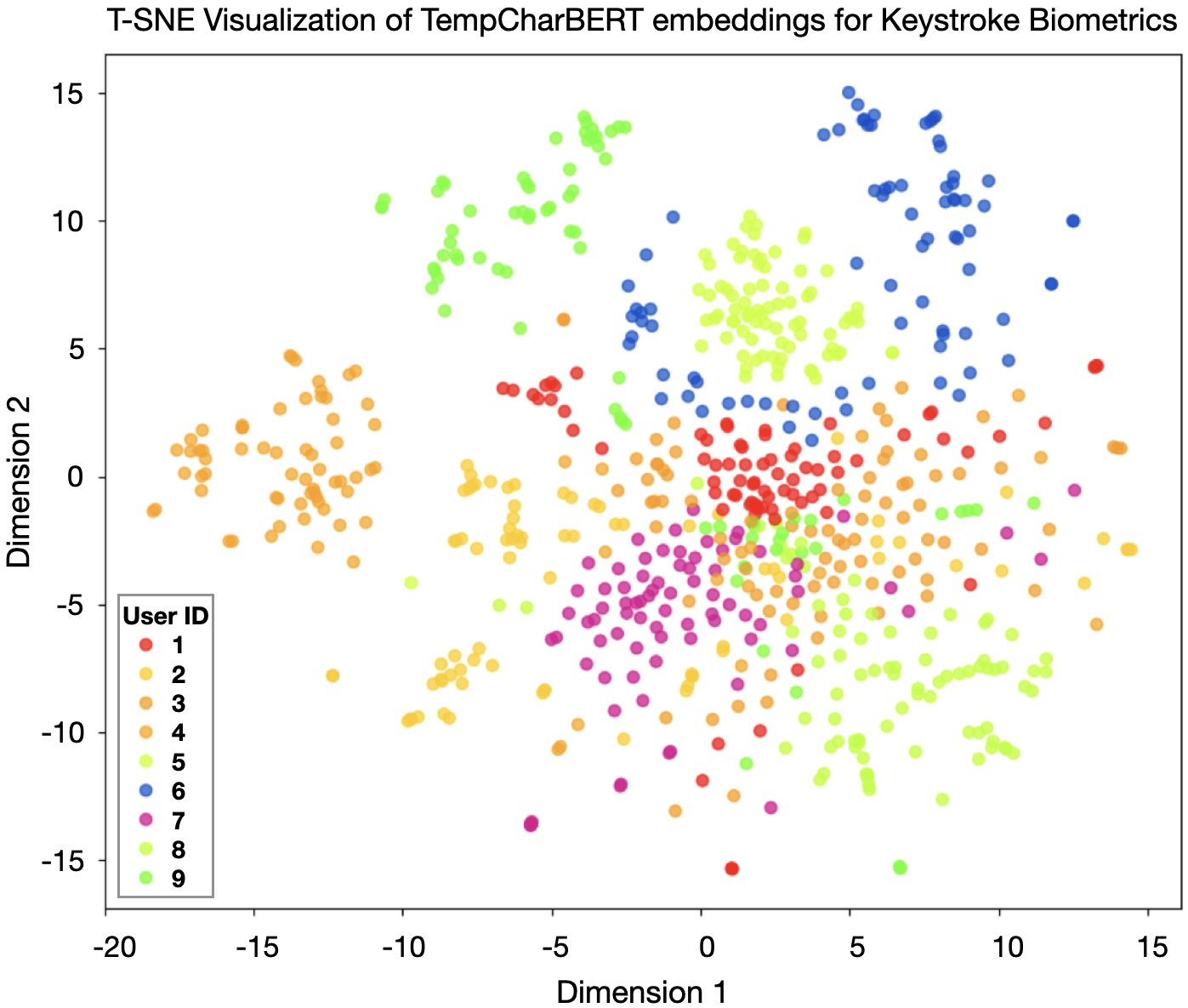} }}%
    \caption{T-SNE visualization of CharBERT embeddings and TempCharBERT embeddings . Temporal keystroke information helps to capture user biometrics.}%
    \label{fig:tsne}%
\end{figure*}

Figure \ref{fig:tsne} illustrates sentences typed by 9 users and projected into a two-dimensional embedding space. Each dot indicates embeddings extracted by the CharBERT and by TempCharBERT embedding layers (i.e.,  $h^i(x)$ and $\overline{h^i}(x)$). Note that while the embeddings from CharBERT carry out only toke-level information, thus neglecting any keystroke temporal information, the embeddings from TempChartBERT are able to capture the user unique typing biometrics. This results in the clusterization of the projected dots by user, suggesting that suitability of the proposed method for keystroke dynamics. 

\section{Experimental Setup}
The proposed model is evaluated on free-text keystroke dynamics. In this section, we first present the dataset used, figure of merit and then a description of the experimental settings.

\subsection{Dataset} 
In this work, we use the dataset described in \cite{gonzalez2023dataset}. It was developed to facilitate research on keystroke dynamic, specially those aimed at distinguishing authentic from fraudulent users. As a free-text dataset, it focus on the evaluation of liveness detection for keystroke dynamics, being well-suited for developing behavioral intrusion detection systems. Although the authors of the dataset used five different methods to create artificial user profiles that were synthetically generated, we only use the keystroke dynamics data from the real users. The original data contains User ID, key code, hold time and flight time, both in milliseconds.The LSIA dataset contains 137 users, and the number of samples per user varies between four to one hundred samples. For testing purposes we separate 20\% of the data. The reader is referred to \cite{gonzalez2023dataset} for more details on the dataset used here.





\begin{table}
    \caption{Results on user identification.}
    \label{tab:identification}
    \centering
    \begin{tabular}{cc}
        \hline
        \textbf{Model} & \textbf{Accuracy} \\\hline
        CharBert & 59.26\% \\
        TempCharBERT & \textbf{90.14\%} \\
        LSTM & 87.50\% \\
        SVM & 61.10\% \\
        Manhattan Distance & 25.74\% \\\hline
        LSTM$_{CharBERT}$ & 50.62\%\\
        LSTM$_{TempCharBERT}$ & \textbf{93.09\%} \\\hline
    \end{tabular}
\end{table}

\subsection{Baselines}
Since our objective is to develop a keystroke biometric system based on PLMs, and considering that there is no evidence of similar work in the literature, CharBERT becomes our main baseline. The closest work to ours is the one presented in \cite{stragapede2023mobile}, which is based on Transformers. Notwithstanding, it lacks the benefits of pre-training and fine-tuning a language model. Moreover, despite our attempt to include it as baseline, their code was not made available by the authors. Three other models are considered as baseline. We chose to include a Long Short-Term Memory (LSTM) model as baseline giving its capability to model sequential temporal information. The LSTM was tested in three different inputs, including CharBERT embeddings, referred to as LSTM$_{CharBERT}$, and TempCharBERT embeddings, namely LSTM$_{TempCharBERT}$. The other two baselines are a Support Vector Machine (SVM) model and the Manhattan Distance.      


\subsection{Figure of Merit}

In our experiments, the proposed model is evaluated on two tasks: user authentication and user identification. In the former, the keystroke systems are required to validate if a claimed identity matches the user template attained during the enrollment phase. For that, performance evaluation is based on EER. Commonly used by biometric systems, the metric represents the point where the False Acceptance Rate (FAR) and the False Rejection Rate are the same (FRR) for a varying threshold. For identification, we use accuracy as to evaluate the model performance.

\subsection{Experimental Settings} 

In our experiments, the proposed model was trained with a batch size 8, using the ADAM optimizer and with learning rate set to 0.00005. We use the BERT base model with 12 layers, 768 hidden units and 12 heads. Fine-tuning took 5 epochs. For the Federated learning experiments, we split the data between 100 clients with the sample ratio set to 0.1.



\begin{table}
    \caption{Results on user authentication.}
    \label{tab:authentication}
    \centering
    \begin{tabular}{cccc}
        \hline
        \textbf{Model} & \textbf{EER} \\\hline
        CharBert & 0.0781\\
        TempCharBERT & \textbf{0.0022}\\
        LSTM & 0.0498\\
        SVM & 0.0822\\
        Manhattan & 0.3337\\\hline
    \end{tabular}
\end{table}

\section{Experimental Results}
In this section, we present three experiments used to evaluate the proposed model. We first discuss the results for the user identification in section 5.1. Next, in section 5.2, the performance of our model for user authentication is also presented. In section 5.3, the performance of our model on a Federated Learning settings is presented. 

\subsection{Experiment I: User Identification} 

Table \ref{tab:identification} presents a comparative analysis between TempCharBERT and the baseline models presented in section 4.2. This comparison underlines the crucial role of temporal data in enhancing CharBert's performance for user identification. Note that incorporating the temporal-character information into hte CharBert architecture is essential to capture biometric information from the user. Without the proposed change to the embedding layer, CharBERT provides the very low accuracy of 59.26 \%, as shown in Table \ref{tab:identification}, while the proposed model achieves accuracy as high as 90.14\%, followed by the LSTM, which achieves 87.50 \%. The statistical model based on the Manhattan Distance exhibits the lowest rates, registering a mere 25.74\% accuracy. The Support Vector Machine (SVM), on the other hand, display better results, 61.10\%, but yet far from the best performance.

\begin{table}
    \caption{CharBERT and TempCharBERT trained for user identification under the FL regime.}
    \label{tab:fl}
    \centering
    \begin{tabular}{ccc}
        \hline
        \textbf{Model} & \textbf{Accuracy} \\\hline CharBert & 61,43\%\\
        TempCharBERT & \textbf{88,16\%}\\\hline
    \end{tabular}
\end{table}

We also validated the embeddings from TempCharBERT on other deep neural network architectures. As seen in Table \ref{tab:identification}, the LSTM$_{TempCharBERT}$ can further improve the results seen with the TempCharBERT. This collaborates our assumption that the temporal-character information can be useful in other architectures. The reason the LSTM$_{TempCharBERT}$ can surpass the performance of TempCharBERT relates to the fact that the recurrent neural network is optimized solely for the keystroke dynamics, while the BERT-like architectures are trained to provide general purpose embeddings that become specialized on the downstream task after fine-tuning. Interestingly, when the CharBERT embeddings are used as input to the LSTM, results are still way bellow expected. Overall, the proposed embeddings led to an increase of 30\% over the CharBERT architecture and a remarkable improvement of 40\% over the LSTM architecture.

\subsection{Experiment II: User Authentication} 

The second experiment focuses on user authentication. Notably, our model outperforms the baselines significantly in terms of EER. Results follow the same trend observed in the first experiments, with TempCharBERT outperforming the baseline solutions. As shown in Table \ref{tab:authentication}, the solution based on the Manhattan distance offers the lowest performance, while. Compared to the SVM and LSTM solutions, the proposed model exhibits a substantial advantage, achieving 0.0022 EER against 0.0822 and 0.0498 achieved by the SVM and LSTM models, respectively. 

\subsection{Experiment III: Federated Learning Settings} 

In this experiment, we investigate the feasibility of training the proposed TempCharBERT on Federated Learning settings. This experiment aims at increasing security while fostering user data privacy at the same time. To achieve this, the training process was decentralized for the user identification task. Local data, containing user typing information, was used to train local models, which was aggregated into the global model at each round using the FEDAvg method. The training was performed in five rounds and ten users were selected per round. Table \ref{tab:fl} shows the results for CharBERT and TempCharBERT with a slight variation of performance compared to the results achieved in the first experiment (see Table \ref{tab:identification}). This show the feasibility of training these models taking privacy into consideration.

\section{Conclusion}

In this work, we addressed  the use of pre-trained language models for keystroke dynamics to enhance user authentication and enable continuous access control. We propose TempCharBERT, which is a variation of the CharBERT architecture that incorporates temporal-character information into the embedding layer. Specifically, we relied on users' hold time and flight time and the model was evaluated on two tasks: user identification and user authentication. Results corroborate to our initial hypothesis, i.e., that more discriminate information is attained by the proposed model which outperformed all the baselines. The embeddings generated with TempCharBERT were also useful on other deep neural network architectures, such as the Long-Short term Memory (LSTM). Finally, to foster user data privacy, we successfully trained TempCharBERT within the Federated Learning settings, and results were as high as the ones achieved using the centralized training setting. This shows the feasibility of training the proposed model while protecting the user privacy.

\bibliographystyle{unsrt}
\bibliography{tempchar}

\end{document}